\documentclass[lettersize,journal]{IEEEtran}
\usepackage{amsmath,amsfonts}
\usepackage{amssymb}
\usepackage{amsthm}
\usepackage{algorithmic}
\usepackage{algorithm}
\usepackage{array}
\usepackage[caption=false,font=normalsize,labelfont=sf,textfont=sf]{subfig}
\usepackage{textcomp}
\usepackage{stfloats}
\usepackage{url}
\usepackage{verbatim}
\usepackage{graphicx}
\usepackage{amsthm}

\usepackage{cite}
\usepackage{color,xcolor}
\usepackage{hyperref}
\usepackage{orcidlink}

\usepackage{todonotes}

\begin{document}

\title{From Regression to Prior-Aware Inference: Solving the ILWE Family in Randomness Leakage Attacks against ML-DSA}

\author{Peiheng Zhang$^{\orcidlink{0009-0006-8831-7679}}$, 
	Yuejun Liu$^{\orcidlink{0000-0002-3821-4050}}$,~\IEEEmembership{Member,~IEEE}, 
	Wei Cheng$^{\orcidlink{0000-0001-9433-7576}}$,~\IEEEmembership{Member,~IEEE}, \\
	Muye Li, 
	Honglin Shao$^{\orcidlink{0009-0009-3944-9968}}$, 
	Yongbin Zhou$^{\orcidlink{0000-0002-1337-0655}}$,~\IEEEmembership{Member,~IEEE} 
	\vspace{-0.6cm}
\thanks{Peiheng zhang, Yuejun Liu, Wei Cheng, Muye Li, Honglin Shao, and Yongbin Zhou are with the School of Cyber Science and Engineering, Nanjing University of Science and Technology, Nanjing 210094, China (e-mail: \{zhangpeiheng, liuyuejun, wei.cheng, limuye, shaohonglin, zhouyongbin\}@njust.edu.cn). Corresponding authors: Yuejun Liu and Yongbin Zhou.}

\thanks{This is a preprint version. This work has been submitted to the IEEE for possible publication.}
}

\markboth{}%
{Shell \MakeLowercase{\textit{et al.}}: A Sample Article Using IEEEtran.cls for IEEE Journals}

\maketitle

\begin{abstract}
ML-DSA is a representative lattice-based signature scheme in post-quantum cryptography standardized by NIST. It relies on signing randomness and rejection sampling to ensure that released signatures are statistically independent of the secret key. Practical implementations, however, may leak partial information about this randomness, and such leakage can transform public signatures into Integer Learning with Errors (ILWE)-type problems, resulting in secret key disclosure risks.

Such randomness leakage attack can be formulated as a two-stage key-recovery procedure, in which leaked partial information and public signatures are first transformed into an ILWE-family instance, and then a recovery solver is applied to recover the secret key. Existing work has mainly focused on the first stage by constructing such instances under different leakage models. By contrast, the role of solver in the subsequent instance-solving stage remains under-explored, 
and existing attacks often rely on ad-hoc model-specific solvers. 
To address this gap, we 
propose a unified framework to systematically evaluate different recovery solvers on leakage-derived ILWE-family instances. 
The framework covers three ILWE instances, including the ordinary ILWE,   
Fiat-Shamir ILWE (FS-ILWE) 
and Concealed ILWE (CILWE) under different scenarios.

Within our {framework}, we explore three classes of solvers. Specifically, the first class is the least-squares regression solvers using ordinary least squares (OLS) and likelihood-based $\ell_2$-estimators. The second class is the robust regression solvers with Huber regression and Cauchy regression. The last class is the prior-aware discrete-inference solvers by using belief propagation (BP), objective-induced greedy search, and bounded-error hill-climbing.
Our experiments show that the solver has a significant impact on the secret-key recovery efficiency.  
In particular, on FS-ILWE, prior-aware discrete-inference reduces the number of informative relations 
by one to two orders of magnitude compared to the baselines: (i) compared with OLS, BP constitutes a reduction by a factor of \(15.4\times\)-\(64.9\times\) in noise-free settings, and by a factor of \(10.5\times\)-\(73.9\times\) in noisy settings. (ii) compared with hill-climbing, BP reduces the number of required relations by a factor of up to \(7.6\times\).
On CILWE, BP is more effective 
with the concealment rate lower than 0.9, while Cauchy regression is more stable for higher rates.
Overall, this work provides a systematic evaluation on different solvers in randomness leakage attacks, and presents new benchmarks for future analysis on ML-DSA.
\end{abstract}

\begin{IEEEkeywords}
Post-Quantum Cryptography, ML-DSA, Integer Learning with Errors, Randomness Leakage Attack.
\end{IEEEkeywords}

\section{Introduction}
\IEEEPARstart{M}{L-DSA}, formerly known as CRYSTALS-Dilithium, was selected by the National Institute of Standards and Technology (NIST) as the primary post-quantum digital signature standard and published as FIPS~204~\cite{fips204} in August 2024. The theoretical security of ML-DSA relies on module-lattice assumptions such as Module-LWE~\cite{langlois2015worst} and Short Integer Solution (SIS)~\cite{ducas2023finding}, while its concrete implementation security also depends on whether intermediate signing values are protected against leakage. ML-DSA follows the Fiat-Shamir-with-aborts paradigm. In each signing attempt, a randomness vector \(\mathbf{y}\) is sampled, a sparse challenge \(\mathbf{c}\) is derived, and the response component is formed as \(\mathbf{z}=\mathbf{y}+\mathbf{c}\mathbf{s}_1\). Rejection sampling is intended to decorrelate the released response \(\mathbf{z}\) from the secret component \(\mathbf{s}_1\), so that \(\mathbf{z}\) is independent of the secret key \(\mathbf{s}_1\) in the absence of leakage on the randomness \(\mathbf y\).

This paradigm no longer holds once an implementation leaks partial information about \(\mathbf{y}\).
Such leakage may arise from power~\cite{qiao2022practical,qiao2024single} or micro-architectural side channels~\cite{damm2025solving}. Together with public signatures, the leaked randomness information enables the construction of Integer Learning with Errors (ILWE)-type problems, which may lead to secret-key recovery. In basic form, ILWE problem can be viewed as an integer linear system \(\mathbf{b}=\mathbf{A}\mathbf{s}+\mathbf{e}\), where the goal is to recover the unknown vector \(\mathbf{s}\). Subsequent randomness leakage attacks on ML-DSA instantiate this core structure in different ways, depending on whether the relations come from exact or noisy randomness-bit leakage~\cite{liu2020security,damm2025one,damm2026less,schubert2026descent,bashiri2026exploiting}, leakage on zero coefficients of the randomness~\cite{damm2025solving}.

Randomness leakage attacks on ML-DSA, therefore, can be viewed as a two-stage key-recovery procedure. The first stage constructs an ILWE-family instance from the leaked randomness information and public signatures, while the second stage applies a recovery solver to obtain the secret key. Recent work has mainly advanced the first stage under different randomness leakage models: randomness-bit leakage gives Fiat-shamir ILWE (FS-ILWE) instance~\cite{liu2020security}, and Concealed ILWE (CILWE) instance~\cite{damm2025solving} from profiling-induced concealed samples in the leakage on zero randomness coefficients. However, the role of solvers in determining the effectiveness of randomness leakage attacks remains under-explored. Existing attacks often adopt model-specific solvers, such as ordinary least squares (OLS) for randomness-bit leakage or robust regression for concealed samples, and these solver choices are not systematically evaluated under a unified formulation. Consequently, it remains unclear whether the attack cost is mainly determined by the first-stage instance construction or by the choice and recovery capability of the solver.

\begin{figure*}[!t]
    \centering
    \includegraphics[width=0.9\textwidth]{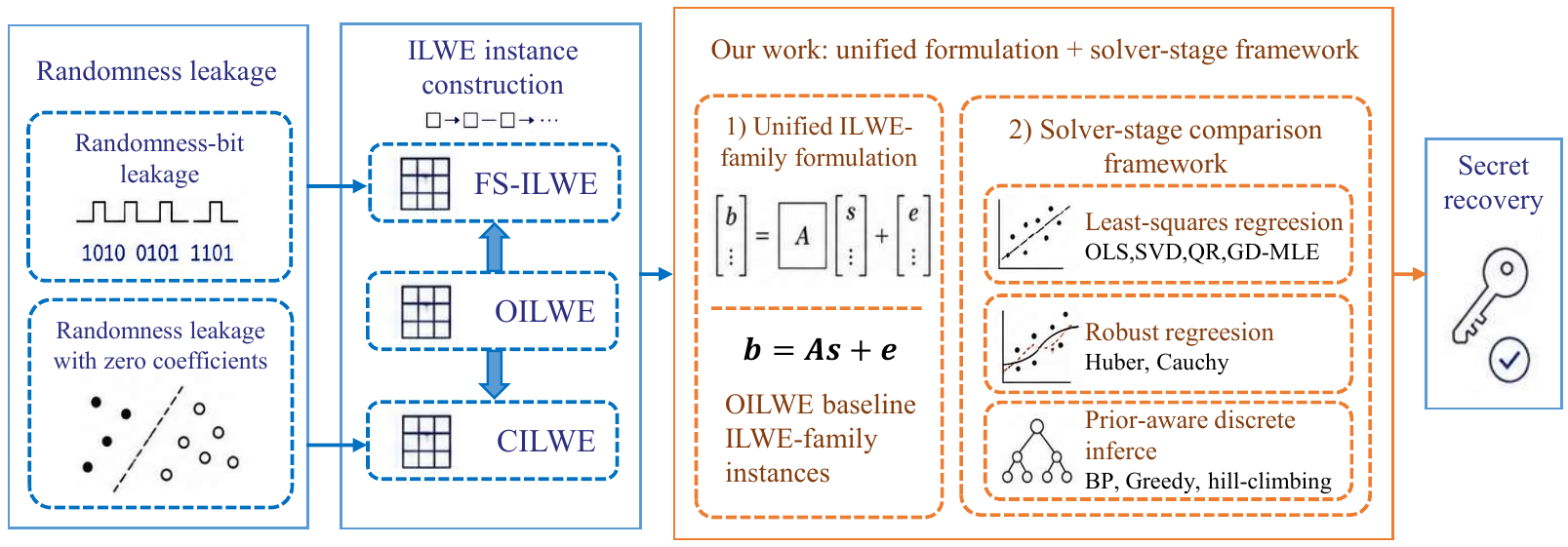}
    \caption{Unified view of randomness-leakage attacks on ML-DSA and the scope of this work.}
    \label{fig:framework}
\end{figure*}

This paper focuses on the instance-solving stage of randomness leakage attacks. Rather than proposing a new leakage-extraction or instance-generation technique, we take the algebraic relations induced by existing randomness leakage models as input and study how different solvers affect recovery on the resulting ILWE-family instances. Fig.~\ref{fig:framework} summarizes this unified view of randomness-leakage attacks on ML-DSA and highlights the scope of this work.

\subsection{Our Contributions}\label{subsec:contributions}

Our contributions are as follows.

\begin{itemize}
\item We propose a unified framework to systematically evaluate differernt solvers on the ILWE-type problems arising in randomness leakage attacks on ML-DSA.
Under this framework, we use OILWE as a baseline and formulate FS-ILWE and CILWE as leakage-derived instances of the same ILWE-type problem, enabling a unified comparison of different solvers. 
We compare three classes of solvers: least-squares regression solvers, robust regression solvers, and prior-aware discrete-inference solvers. Specifically, least-squares regression solvers include ordinary least squares and likelihood-based $\ell_2$-estimators; robust regression solvers include Huber regression, and Cauchy regression; and prior-aware discrete-inference solvers include belief propagation (BP), objective-induced greedy, and bounded-error hill-climbing. We show that least-squares and robust regression solvers both rely on real-valued estimation followed by coordinate-wise rounding, whereas prior-aware discrete-inference solvers search or infer directly over the finite secret space. For each ILWE-family instance, we evaluate all solvers\footnote{
\textcolor{black}{For the solvers, we implemented OLS and the likelihood-based \(\ell_2\) estimator according to~\cite{bootle2018lwe,zhang2024solving}, and instantiated Greedy-\(\ell_2\) as a bounded coordinate-wise local-search solver by following the framework~\cite[Alg.~7]{ravi2024defeating} (we adapted the score function to be the reduction in the squared residual norm). We used the robust-regression implementations from~\cite{damm2025solving}, the BP implementation from~\cite{hermelink2025generic}, and the hill-climbing implementation from~\cite{schubert2026descent}.}} in the three classes under the same problem formulation. In the ordinary ILWE baseline, BP reduces the number of required samples by a factor of up to \(56.1\times\) compared to OLS.

\item We apply this framework to FS-ILWE induced by single-bit randomness leakage in ML-DSA. Our experiments show that prior-aware discrete-inference solvers reduce the number of informative relations by one to two orders of magnitude compared to the prior work: On noise-free FS-ILWE, BP reduces the number of required informative relations by a factor of \(15.4\times\)-\(64.9\times\) compared to OLS across the three ML-DSA security levels. On noisy FS-ILWE, for noise levels at which OLS remains recoverable, BP reduces the number of required informative relations by a factor of up to \(73.9\times\). 
Beyond improving relation efficiency, prior-aware discrete-inference solvers extend the recoverable range to FS-ILWE instances induced by lower randomness leakage bit positions.
Within the class of prior-aware discrete solvers, BP reduces the number of required informative relations by a factor of up to \(7.6\times\) compared to hill-climbing.

\item We also apply the framework to CILWE arising from leakage on zero coefficients of the randomness. In this instance, the comparison between robust regression solvers and prior-aware discrete-inference solvers is regime dependent.
BP achieves better recovery performance than Cauchy regression, the strongest robust-regression baseline in our experiments, for concealment rates below \(0.9\). In contrast, Cauchy regression becomes more effective when the concealment rate reaches \(0.9\) or higher.
These results identify the concealment-rate regime in which BP provides a clear advantage, while also showing that Cauchy regression can become preferable when the informative samples are extremely sparse.
\end{itemize}

\subsection{Related Work} \label{subsec1.1}

\paragraph{OILWE and regression-based solvers}
The Integer Learning with Errors (ILWE) problem, an LWE variant without modular reduction, was formally introduced and systematically studied by Bootle et al.~\cite{bootle2018lwe}. Given an ILWE instance $\mathbf{b}=\mathbf{A}\mathbf{s}+\mathbf{e}$, they showed that ordinary least squares can compute a real-valued approximation $\hat{\mathbf{s}}$ of $\mathbf{s}$, and that coordinate-wise rounding recovers $\mathbf{s}$ when sufficiently many samples are available under standard distributional assumptions.
Beyond the abstract ILWE formulation, prior work~\cite{bootle2018lwe,barthe2019galactics} has shown that side-channel attacks on BLISS and its variant GALACTICS can be reduced to ILWE-type problems and solved using regression-based techniques. More recently, Zhang et al.~\cite{zhang2024solving} revisited ILWE solving from a numerical optimization perspective. They observed that OLS via normal equations incurs high computational cost due to matrix inversion at large sample sizes and proposed a batch gradient descent variant to improve efficiency in BLISS-related settings. 
Overall, these results establish ILWE as a unifying abstraction for leakage-induced linear recovery in lattice-based signatures. However, existing work primarily focuses on least-squares regression solvers and their numerical optimization, without systematically studying methods that exploit the discrete small-secret structure.

\paragraph{Randomness-bit leakage attacks and FS-ILWE}
Liu et al.~\cite{liu2020security} first formalized randomness-bit leakage attacks against lattice-based Fiat-Shamir signatures, showing that one leaked bit of the signing randomness $\mathbf{y}$ can be combined with public signatures to obtain linear constraints on the secret key. This yields Fiat-Shamir ILWE (FS-ILWE), an ILWE-type problem instance induced by randomness-bit leakage. Subsequent work has refined and extended this basic construction along several directions. Wang et al.~\cite{wang2021integer} generalized ILWE to secret-dependent errors and showed that the resulting variants remain solvable by OLS. Qiao et al.~\cite{qiao2022practical,qiao2024single} demonstrated practical randomness-bit leakage on unprotected and masked Dilithium implementations. Damm et al.~\cite{damm2025one,damm2026less} introduced leakage-position-independent transformations that reduce the number of required informative signatures in high-order bit and noisy bit leakage settings. Schubert et al.~\cite{schubert2026descent} further exploited the bounded-error structure through hill-climbing local search, whereas Bashiri et al.~\cite{bashiri2026exploiting} derived bit-level constraints and solved them using integer linear and nonlinear programming. These works primarily advance the instance construction stage or develop solvers tailored to specific leakage formulations. In contrast, we focus on the second stage and compare multiple classes of solvers for noise-free FS-ILWE and noisy FS-ILWE under a unified ILWE-family framework.

\paragraph{Profiling side-channel attacks and CILWE}
Ulitzsch et al.~\cite{ulitzsch2024profiling} and Berzati et al.~\cite{berzati2023exploiting} showed that profiling-based side-channel attacks on ML-DSA can identify zero coefficients of the randomness $\mathbf{y}$ and transform the resulting observations into linear systems for key recovery, typically solved via integer linear programming (ILP) or linear-algebraic methods. Damm et al.~\cite{damm2025solving} formalized this imperfect classification setting as Concealed ILWE (CILWE), where the collected samples form a mixture of zero-error ILWE relations and misclassified observations, which may behave as zero-knowledge samples. They showed that ordinary least squares becomes ineffective because zero-knowledge samples violate the assumption that the error is centered and independent of the secret-dependent term, while ILP is limited by scalability. To improve robustness, Damm et al. introduced robust regression solvers, including Huber and Cauchy regression, together with iterative reweighted least squares (IRLS). Existing works primarily focus on improving the robustness of specific recovery procedures for such mixed systems, using ILP or robust regression techniques. In contrast, the recovery layer itself has not been systematically compared across solver paradigms under a unified framework.

\paragraph{Other solvers and applications}
\textcolor{black}{
Beyond continuous estimators followed by coordinate-wise rounding, prior-aware discrete-inference solvers have been developed to exploit the bounded discrete secret distribution directly. In fault and decryption-failure attacks against ML-KEM and other lattice-based KEMs, the key recovery problem is converted into a system of linear inequalities, which can be solved using Bayesian updating and BP~\cite{prokop2021fault,hermelink2021fault,delvaux2021roulette}. 
Ravi et al.~\cite{ravi2024defeating} proposed a greedy algorithm for solving inequalities arising from side-channel attacks against ML-KEM. Hermelink et al.~\cite{hermelink2025generic} subsequently generalized these attack-specific approaches into a distribution-hint framework for LWE-based cryptosystems and developed corresponding BP and greedy solvers. More recently, Hermelink et al.~\cite{hermelink2025finding} applied this distribution-hint framework to side-channel leakage of the signing randomness \(\mathbf{y}\) in masked ML-DSA, using BP and a filtering technique to recover the secret key. These works demonstrate the strength of prior-aware inference for side-channel-assisted key recovery in lattice-based cryptography, but their focus is on inequality-derived constraints or distribution-hint attack pipelines. Our work is complementary in that it studies ILWE-family linear recovery problems produced after leakage-to-relation conversion randomness leakage attacks against ML-DSA, and provides a unified comparison of solvers for the resulting recovery systems.}

\begin{table}[!t]
    \centering
    \caption{ILWE-type problems and representative existing solvers}
    \label{tab:ilwe_type_summary}
    \scriptsize
    \renewcommand{\arraystretch}{1.12}
    \setlength{\tabcolsep}{3.6pt}
    \begin{tabular}{l p{0.50\columnwidth} p{0.20\columnwidth}}
        \hline
        \textbf{Problem Type} & 
        \textbf{Existing Recovery Methods} & 
        \textbf{References} \\
        \hline
        OILWE & 
        OLS, likelihood-based estimators. & 
        \cite{bootle2018lwe,zhang2024solving} \\
        \hline
        FS-ILWE & 
        OLS, hill-climbing. & 
        \cite{liu2020security,damm2025one,damm2026less,schubert2026descent} \\
        \hline
        CILWE & 
        OLS, ILP, Huber, Cauchy. & 
        \cite{damm2025solving,ulitzsch2024profiling,berzati2023exploiting} \\
        \hline
    \end{tabular}
\end{table}

\section{Preliminaries}\label{sec2}

In this section, we present notation and definitions.

\subsection{Notations}
For a real number \(r\in\mathbb{R}\), we denote by \(\lfloor r\rceil\) its nearest integer. For two integers \(a\leq b\), we write $[a,b]=\{a,a+1,\ldots,b\}$, and use \([\pm \alpha]\) as shorthand for \([-\alpha,\alpha]\cap\mathbb{Z}\). Vectors are written in bold lowercase letters and matrices in bold uppercase letters. For a vector \(\mathbf{x}=(x_1,\ldots,x_n)^T\in\mathbb{R}^n\), we use $\|\mathbf{x}\|_2\ \text{and}\ \|\mathbf{x}\|_\infty=\max_{1\leq i\leq n}|x_i|$ for the Euclidean norms $(\ell_2)$ and infinity norms $(\ell_\infty)$, respectively. For two vectors \(\mathbf{a},\mathbf{b}\in\mathbb{R}^n\), their inner product is $\langle \mathbf{a},\mathbf{b}\rangle = \sum_{i=1}^{n}a_i b_i = \mathbf{a}^{T}\mathbf{b}$.

For a matrix \(\mathbf{A}\in\mathbb{R}^{m\times n}\), we denote its \(i\)-th row by \(\mathbf{a}_i\) and its \((i,j)\)-th entry by \(a_{i,j}\). For factor-graph or sparse-row solvers, we also write $\mathcal{N}(i)=\{j:a_{i,j}\neq0\},\ \mathcal{M}(j)=\{i:a_{i,j}\neq0\}$, for the variables involved in the \(i\)-th relation and the samples involving the \(j\)-th variable, respectively. We use \(\mathbf{e}_j\) to denote the \(j\)-th standard basis vector. We use \(\mathrm{Ber}(p)\) for a Bernoulli random variable with success probability \(p\), and \(\mathbf{1}[\mathcal{E}]\) for the indicator of an event \(\mathcal{E}\).

Throughout the paper, ML-DSA operations are viewed coefficient-wise over the ring $\mathcal{R}_q=\mathbb{Z}_q[x]/(x^n+1)$, where \(n=256\) for ML-DSA and \(q\) is the modulus. A ring element $a=\sum_{i=0}^{n-1}a_i x^i\in\mathcal{R}_q$ is identified with its coefficient vector $\mathbf{a}=(a_0,a_1,\ldots,a_{n-1})\in\mathbb{Z}_q^n$.

\subsection{ML-DSA} \label{subsec:mldsa}

ML-DSA, formerly known as CRYSTALS-Dilithium, is a lattice-based digital signature scheme standardized by NIST. It follows the Fiat-Shamir-with-aborts paradigm and operates over the polynomial ring $\mathcal{R}$, and its modular reduction $\mathcal{R}_{q}$. Since each element of \(\mathcal{R}\) can be viewed as a \(256\)-dimensional coefficient vector, polynomial multiplication in ML-DSA can be rewritten as structured linear relations over the integers. Table~\ref{tab:mldsa_params} lists the ML-DSA parameters. The pair $(k,\ell)$ specifies the module dimensions, \(\eta\) bounds the coefficients of the secret vectors, \(\gamma_1\) determines the range of the randomness, and \(\tau\) is the Hamming weight of the challenge polynomial. We write $\beta=\eta\tau$, which upper bounds the coefficient growth caused by the product \(\mathbf{c}\mathbf{s}_1\).

\begin{table}[tbp]
    \centering
    \caption{ML-DSA parameters used in ILWE-type recovery experiments}
    \label{tab:mldsa_params}
    \scriptsize
    \renewcommand{\arraystretch}{1.12}
    \setlength{\tabcolsep}{3.6pt}
    \begin{tabular}{lcccccc}
        \hline
        \textbf{Scheme} & \(\boldsymbol{(k,\ell)}\) & \(\boldsymbol{n}\) & \(\boldsymbol{\eta}\) & \(\boldsymbol{\tau}\) & \(\boldsymbol{\beta=\eta\tau}\) & \(\boldsymbol{\gamma_1}\) \\
        \hline
        ML-DSA-44 (L2) & \((4,4)\) & \(256\) & \(2\) & \(39\) & \(78\)  & \(2^{17}\) \\
        ML-DSA-65 (L3) & \((6,5)\) & \(256\) & \(4\) & \(49\) & \(196\) & \(2^{19}\) \\
        ML-DSA-87 (L5) & \((8,7)\) & \(256\) & \(2\) & \(60\) & \(120\) & \(2^{19}\) \\
        \hline
    \end{tabular}
\end{table}

We only recall the part of ML-DSA signing that is relevant to our attacks. Let $\mathbf{s}_1\in \mathbf{S}_\eta^\ell$ be the secret component, where $\mathbf{S}_\eta=\{w\in\mathcal{R}:\|w\|_\infty\leq \eta\}$, and let $\mathbf{y} \in \mathbf{S}_{\gamma_1}^{\ell}$ be the randomness vector sampled during signing. The signer derives a sparse challenge polynomial $\mathbf{c}\in \mathbf{B}_\tau$, \(\mathcal B_\tau\) denotes the set of polynomials with exactly \(\tau\) nonzero coefficients in \(\{\pm1\}\), and computes $\mathbf{z}=\mathbf{y}+\mathbf{c}\mathbf{s}_1$. The signature is released only after rejection sampling. Since $\|\mathbf{c}\mathbf{s}_1\|_\infty\leq \beta$, the shift introduced by \(\mathbf{c}\mathbf{s}_1\) is hidden by the randomness \(\mathbf{y}\) in the leakage-free setting. Hence the public values \((\mathbf{z},\mathbf{c})\) do not directly reveal useful information about \(\mathbf{s}_1\).

\subsection{ILWE, FS-ILWE, and CILWE}\label{subsec2.2}

We consider ILWE-type problems that arise from linear relations with error over the integers. In ordinary ILWE, each sample satisfies $b=\langle \mathbf{a},\mathbf{s}\rangle+e$, where $\mathbf{s}\in\mathbb{Z}^{n}$ is the unknown secret vector, $\mathbf{a}\in\mathbb{Z}^{n}$ is the public sample vector, and $e\in\mathbb{Z}$ is an error term. Given $m$ samples, the stacked form is
\begin{equation}
    \mathbf{b} = \mathbf{A}\mathbf{s} + \mathbf{e}.
    \label{eq:ILWE}
\end{equation}
The goal is to recover $\mathbf{s}$ from $(\mathbf{A},\mathbf{b})$.

FS-ILWE is the ILWE variant arising from randomness-bit leakage in Fiat-Shamir signatures. In ML-DSA, each row of \(\mathbf{c}\) is derived from a sparse challenge polynomial and contains \(\tau\) nonzero entries in \(\{\pm1\}\). After informative-relation extraction, the attacker obtains bounded-error $\beta_{\mathrm{eff}}$ relations
\begin{equation}
    \tilde z_i
    =
    \langle \mathbf{c}_i,\mathbf{s}\rangle
    +
    \tilde y_i,
    \qquad
    |\tilde y_i|\leq \beta_{\mathrm{eff}}.
    \label{eq:FS-ILWE}
\end{equation}
CILWE is a concealed ILWE instance in which only a fraction of the selected samples are truly informative. We write each selected sample in the unified ILWE-type form
\begin{equation}
    \begin{aligned}
    z_i &= \langle\mathbf{c}_i,\mathbf{s}\rangle+y_i,\ 
    \\
    y_i &=
    \begin{cases}
    0, & \text{with probability } 1-p_{\mathrm{con}},\\
    \tilde{y}_i\leftarrow\chi_{con}, & \text{with probability } p_{\mathrm{con}},
    \end{cases}
    \end{aligned}
    \label{eq:CILWE}
\end{equation}
where $p_{\mathrm{con}}$ is the concealment rate and $\chi_{\mathrm{con}}$ denotes the error distribution of concealed samples.

\section{The Ordinary Integer-LWE problem}\label{sec3}

This section focuses on ordinary ILWE, as defined in Section~\ref{subsec2.2}. We first review the OLS solver with coordinate-wise rounding and its sample-complexity bound. We then compare least-squares regression solvers, robust regression solvers, and prior-aware discrete-inference solvers under settings in which the entries of each public sample vector and the error term are independently drawn from prescribed uniform distributions.

\subsection{Review of OILWE and Least-Squares Regression Solvers} \label{subsec3.1}

The ordinary ILWE problem was formally defined and systematically studied by Bootle et al.~\cite{bootle2018lwe} as an LWE variant without modular reduction. As in Eq.~\eqref{eq:ILWE}, ordinary ILWE gives linear relations with error over the integers, which allows the problem to be relaxed to real-valued regression under suitable distributional assumptions. The standard baseline is the OLS solver: first compute the objective
\begin{equation}
    \hat{\mathbf{s}}_{\mathrm{LS}}= \arg\min_{\mathbf{x}\in\mathbb{R}^{n}} \|\mathbf{A}\mathbf{x}-\mathbf{b}\|_{2}^{2},
    \label{eq:ordinary_ilwe_ls}
\end{equation}
and then recover the integer secret by coordinate-wise rounding whenever
\begin{equation}
    \|\mathbf{s}-\hat{\mathbf{s}}_{\mathrm{LS}}\|_{\infty}<\frac{1}{2}.
    \label{eq:round-eq}
\end{equation}

For a candidate $\mathbf{x}$, let $r_i(\mathbf{x})=b_i-\langle \mathbf{a}_i,\mathbf{x}\rangle$. Then Eq.~\eqref{eq:ordinary_ilwe_ls} minimizes $\sum_{i=1}^{m} \|r_i(\mathbf{x})\|^2_2$. Normal equations, singular value decomposition pseudoinverse (SVD), QR decomposition (QR), and fully converged gradient-based $\ell_2$ solvers are therefore different numerical realizations of the same continuous least-squares estimator. Under a subgaussian error model, the same objective also follows from maximum likelihood, hence gradient-descent maximum likelihood estimation (GD-MLE) can be seen as an iterative optimizer for the same $\ell_2$-based estimator rather than a different recovery method~\cite{zhang2024solving}.

The theoretical recovery threshold of these least-squares regression solvers is governed by the rounding condition in Eq.~\eqref{eq:round-eq}. Let \(\sigma_a^2\) denotes the variance of \(X_a\), and \(\tau_a\) is sub-Gaussian parameter; similarly for \(X_e\). Suppose $\mathcal{X}_{a}$ is $\tau_a$-subgaussian, $\mathcal{X}_{e}$ is $\tau_e$-subgaussian, and $\mathbf{A}\leftarrow\mathcal{X}_{a}^{m\times n}$, $\mathbf{e}\leftarrow\mathcal{X}_{e}^{m}$, Bootle et al.~\cite{bootle2018lwe} show that there exist constants $C_1,C_2>0$ such that, for all $\kappa\geq 1$, if
\begin{equation}
    m \geq 4\frac{\tau_a^{4}}{\sigma_a^{4}}(C_1 n+C_2\kappa)\ \text{and}\ m \geq 32\frac{\tau_e^{2}}{\sigma_a^{2}}\log(2n),
    \label{eq:ILWE_ls_minsample}
\end{equation}
then Eq.~\eqref{eq:round-eq} holds, and thus $\lfloor \hat{\mathbf{s}}_{\mathrm{LS}}\rceil=\mathbf{s}$ with probability at least $1-\frac{1}{2n}-2^{-\kappa}$. The first condition ensures sufficient conditioning of the sample matrix, while the second controls the error perturbation. Consequently, OLS, SVD, QR, GD-MLE, and other fully converged $\ell_2$-based regression solvers share the same sample-complexity boundary. Thus, these solvers implement the same \(\ell_2\)-based estimation criterion through different numerical decompositions or optimization procedures, rather than changing the underlying recovery objective.

\subsection{Solving OILWE via Robust Regression Solvers} \label{subsec3.2}

Least-squares regression is well matched to ordinary ILWE when the errors are centered, independent, and well behaved, but it is sensitive to large residuals because of the $\ell_2$ loss. Robust regression solvers replace the $\ell_2$ loss with robust loss functions that reduce the influence of outlying samples. For residuals $r_i(\mathbf{x})$, a general robust estimator is
\begin{equation}
    \hat{\mathbf{s}}_{\rho} = \arg\min_{\mathbf{x}\in\mathbb{R}^{n}} \sum_{i=1}^{m}\rho(r_i(\mathbf{x})),
    \label{eq:ordinary_ilwe_robust}
\end{equation}
where $\rho$ is a robust loss. Since the optimization is still over $\mathbb{R}^n$, the resulting estimate must still be rounded coordinate-wise for exact integer recovery.

We consider three representative robust regression solvers. The $\ell_1$ loss uses $\rho(r)=|r|$ and can be solved through a linear-programming formulation. The Huber loss is quadratic near zero and linear for large residuals,
\begin{equation*}
    \rho_{\delta}(r)=
    \begin{cases}
        \frac{1}{2}r^2, & |r|\leq\delta,\\
        \delta\left(|r|-\frac{1}{2}\delta\right), & |r|>\delta,
    \end{cases}
\end{equation*}
where $\delta>0$ controls the transition point. The Cauchy loss, $\rho_{\mathrm{Cauchy}}(r) = \frac{c^2}{2}\log\left(1+\left(\frac{r}{c}\right)^2\right)$, downweights large residuals more aggressively, but leads to a generally non-convex objective. Huber and Cauchy regression solvers are implemented by iterative reweighted least squares: at iteration $t$, residuals $r_i^{(t)}$ define weights $w_i^{(t)}=\rho'(r_i^{(t)})/r_i^{(t)}$,\footnote{$\rho'(\cdot)$ denotes the first derivative of the loss function.} and the next estimate is obtained by solving the weighted least-squares problem $\mathbf{x}^{(t+1)} = \arg\min_{\mathbf{x}\in\mathbb{R}^{n}} (\mathbf{b}-\mathbf{A}\mathbf{x})^{T} \mathbf{W}^{(t)} (\mathbf{b}-\mathbf{A}\mathbf{x})$.

From the sample-complexity viewpoint, robust regression solvers should be viewed as a robust continuous-estimation baseline rather than a theoretically stronger solver for ordinary ILWE. Under this model, OLS can exploit all available samples through the $\ell_2$-loss objective and is therefore well matched to the recovery problem. Robust regression solvers are designed to downweight samples with large residuals. Hence, robust regression is not expected to reduce the theoretical sample requirement relative to OLS. Their advantage appears in non-ideal leakage-induced instances, such as noisy FS-ILWE or CILWE, where large residuals, concealed samples, or leakage-induced outliers can dominate an $\ell_2$ estimator. Nevertheless, these solvers remain continuous regression and rely on coordinate-wise rounding for exact integer recovery.

\subsection{Solving OILWE via Prior-Aware Discrete-Inference Solvers} \label{subsec3.3}

The solvers discussed above are continuous regression: they optimize over \(\mathbb{R}^n\), output a real-valued estimate \(\hat{\mathbf{s}}\), and require coordinate-wise rounding for exact integer recovery. In ILWE-based cryptanalytic settings, however, the secret usually has a known small-integer prior. In ML-DSA, the secret key \(\mathbf{s}\in\mathcal{S}\subseteq\{-\eta,\ldots,\eta\}^n\). Prior-aware discrete-inference solvers incorporate this structural prior by formulating recovery directly over the bounded integer candidate space \(\mathcal S\). Consequently, they output an integer-valued candidate, rather than relying on a continuous estimate followed by coordinate-wise rounding.

We consider three representative prior-aware discrete-inference solvers. BP performs probabilistic inference over the possible coefficient values. Objective-induced greedy searches over integer candidates by selecting local updates that decrease a residual objective. We also include bounded-error hill-climbing, which evaluates candidates by their consistency with an effective bounded-error region.

\subsubsection{Belief Propagation Solver}

Belief propagation (BP) is a prior-aware discrete inference that treats each secret coefficient as a discrete random variable. In contrast to continuous regression solvers, BP does not first compute a real-valued estimate and then apply coordinate-wise rounding. Instead, it directly compares candidate values in the finite secret space and outputs the most likely integer secret.

Consider the ordinary ILWE relation in~\eqref{eq:ILWE}, where \(s_j\in\mathcal{S}_j\) and the error distribution is \(\chi_e\). For a candidate secret \(\mathbf{x}\), the residual of the \(i\)-th sample is \(r_i(\mathbf{x})\). Hence, the likelihood of this sample is determined by \(\chi_e(r_i(\mathbf{x}))\). Let \(\mathcal{N}(i)\) denote the set of secret variables involved in the \(i\)-th relation. The corresponding factor is
\[
    \phi_i(\mathbf{x}_{\mathcal{N}(i)})
    =
    \chi_e\!\left(
        b_i-\sum_{j\in\mathcal{N}(i)} a_{i,j}x_j
    \right).
\]
Combining all sample likelihoods with the coefficient-wise secret prior gives
\begin{equation}
    \Pr[\mathbf{x}\mid\mathbf{A},\mathbf{b}]
    \propto
    \prod_{j=1}^{n}\Pr[x_j]
    \prod_{i=1}^{m}
    \phi_i(\mathbf{x}_{\mathcal{N}(i)}).
    \label{eq:bp_prior}
\end{equation}

This factorization defines a factor graph whose variable nodes are the secret coefficients and whose factor nodes are the ILWE samples. BP approximates the marginal posterior of each coefficient by iteratively exchanging messages on this graph. For \(\mathcal{M}(j)\), the variable-to-factor message is
\begin{equation}
    \nu_{j\to i}^{(t)}(v)
    \propto
    \Pr[x_j=v]
    \prod_{i'\in\mathcal{M}(j)\setminus\{i\}}
    \mu_{i'\to j}^{(t-1)}(v),
    \quad v\in\mathcal{S}_j ,
    \label{eq:v-t-f}
\end{equation}
and the factor-to-variable message is
\begin{equation}
    \begin{aligned}
    \mu_{i\to j}^{(t)}(v)
    \propto
    \sum_{\substack{x_{\ell}\in\mathcal{S}_{\ell}\\
    \ell\in\mathcal{N}(i)\setminus\{j\}}}
    &
    \chi_e\!\left(
        b_i-a_{i,j}v
        -\sum_{\ell\in\mathcal{N}(i)\setminus\{j\}}
        a_{i,\ell}x_{\ell}
    \right)
    \\
    &\cdot
    \prod_{\ell\in\mathcal{N}(i)\setminus\{j\}}
    \nu_{\ell\to i}^{(t)}(x_{\ell}),
    \quad v\in\mathcal{S}_j .
    \end{aligned}
    \label{eq:f-t-v}
\end{equation}

In implementation, we do not enumerate all assignments in \(\mathcal N(i)\setminus\{j\}\) when computing (10). Since both the coefficients \(a_{i,\ell}\) and the secret coefficients \(x_\ell\) take bounded integer values in our ILWE-family instances, the neighboring contribution
\[
    \sum_{\ell\in \mathcal N(i)\setminus\{j\}} a_{i,\ell}x_\ell
\]
has bounded integer support. We compute the factor-to-variable messages by dynamic programming over this contribution sum, equivalently by repeated convolution of the incoming contribution distributions. Therefore, the update cost is polynomial in the row weight and in the residual-support width, rather than exponential in \(|\mathcal N(i)|\).

After \(T\) iterations, the approximate marginal belief of \(x_j\) is
\[
    q_j(v)
    \propto
    \Pr[x_j=v]
    \prod_{i\in\mathcal{M}(j)}
    \mu_{i\to j}^{(T)}(v),
    \quad v\in\mathcal{S}_j .
\]
The recovered coefficient is then chosen by the maximum-a-posteriori rule $\hat{s}_j=\arg\max_{v\in\mathcal{S}_j} q_j(v)$. Thus, BP directly outputs an integer candidate \(\hat{\mathbf{s}}\in\mathcal{S}_1\times\cdots\times\mathcal{S}_n\), without requiring coordinate-wise rounding.

The reason BP can solve ordinary ILWE is that each relation gives a likelihood score for every small-integer candidate, and these likelihoods accumulate over many samples. If the error distribution is not too wide and the number of relations is large enough, the posterior mass of the true secret becomes separated from that of wrong candidates. BP succeeds when the available samples create a clear posterior separation between the true bounded integer secret and all competing candidates.

Conversely, BP fails when such posterior separation cannot be formed. This may happen if the number of samples is insufficient, if the error distribution is too wide, if the relations are too weak to distinguish nearby integer candidates, or if the assumed likelihood model does not match the actual error distribution. Since BP is an approximate inference solver on a loopy factor graph, dense factors, many weak relations, or extremely small likelihoods may also cause slow convergence or numerical instability. In these cases, the marginals remain close to the prior, converge to an incorrect posterior, or fail to stabilize, and BP can no longer reliably recover the secret.

\subsubsection{Objective-induced Greedy Solver} We next instantiate a prior-aware greedy solver from the residual objectives introduced above. Unlike BP, which maintains posterior distributions over coefficient values, the greedy solver keeps a single integer candidate \(\mathbf{x}^{(t)}\in\mathcal{S}\) and improves it through local updates. The key idea is to transform the objective functions of continuous solvers into scores for discrete candidate updates.

For a candidate \(\mathbf{x}\in\mathcal{S}\), define the residual loss $\mathcal{L}_{\rho}(\mathbf{x}) = \sum_{i=1}^{m} \rho \left(b_i-\langle \mathbf{a}_i,\mathbf{x}\rangle\right)$, where \(\rho\) is one of the residual losses introduced in the previous subsection. The $\ell_2$ loss gives an OLS-induced greedy solver, while the \(\ell_1\), Huber, and Cauchy losses give robust-induced greedy solvers.

At iteration \(t\), the solver considers admissible coordinate updates $\mathbf{x}^{(t)} \longmapsto \mathbf{x}^{(t)}+\Delta\mathbf{e}_j \in\mathcal{S}$, where \(\mathbf{e}_j\) is the \(j\)-th standard basis vector and \(\Delta\) is an integer change. The score of this update is defined as the decrease of the residual loss:
\[
    \mathsf{Score}_{\rho}^{(t)}(j,\Delta) = \mathcal{L}_{\rho}(\mathbf{x}^{(t)}) - \mathcal{L}_{\rho} \left(\mathbf{x}^{(t)}+\Delta\mathbf{e}_j \right).
\]
Thus, a larger score means that the update decreases the chosen residual objective more significantly. The solver selects the admissible update with the largest positive score, $(j^{\ast},\Delta^{\ast}) = \arg\max_{\substack{j,\Delta \mathbf{x}^{(t)}+\Delta\mathbf{e}_j\in\mathcal{S}}} \mathsf{Score}_{\rho}^{(t)}(j,\Delta)$, and applies it as $\mathbf{x}^{(t+1)} = \mathbf{x}^{(t)} + \Delta^{\ast}\mathbf{e}_{j^{\ast}}$, provided that the best score is positive. Otherwise, the iteration stops, or a prescribed restart rule is applied.

This construction turns continuous residual objectives into prior-aware discrete search criteria. With the $\ell_2$ loss, the greedy solver chooses the integer update that most decreases the OLS objective. Hence, the greedy variants search directly over the bounded integer space \(\mathcal{S}\), avoiding the coordinate-wise rounding requirement of regression solvers. The limitation is that they remain local-search methods and may depend on initialization, update order, and restart strategies.

Hill-climbing is used as a bounded-error consistency-based local-search solver over the integer candidate space. Similar to the greedy solvers, it maintains a candidate \(\mathbf{x}\in\mathcal{S}\) and iteratively applies local modifications. However, its score is not defined by the decrease of an \(\ell_2\), \(\ell_1\), Huber, or Cauchy residual objective. Instead, it evaluates how far the residuals deviate from an effective error interval. For a residual $r_i(\mathbf{x})$, and an effective bound \(\beta_{\mathrm{eff}}^{\star}\), we use
\[
    \mathsf{Score}_{\mathrm{HC}}(\mathbf{x}) = \sum_{i=1}^{m} \max\{0, |r_i(\mathbf{x})|-\beta_{\mathrm{eff}}^{\star}\}.
\]
Residuals inside \([-\beta_{\mathrm{eff}}^{\star},\beta_{\mathrm{eff}}^{\star}]\) are regarded as consistent and receive no penalty, while only the excess outside this interval contributes to the score. If the error distribution has known bounded support, \(\beta_{\mathrm{eff}}^{\star}\) is chosen from this support; otherwise, it is treated as an empirical effective bound calibrated from the error scale or experiments. Thus, hill-climbing acts as a bounded-error consistency search and differs from objective-induced Greedy mainly in its scoring principle. Detailed algorithmic strategies and scoring variants can be found in~\cite{schubert2026descent}.

\subsubsection{Sample-Complexity Interpretation}

Prior-aware discrete-inference solvers have a different recovery condition from continuous-regression-then-rounding solvers. Continuous regression solvers output a real-valued vector $\hat{\mathbf{s}}\in\mathbb{R}^{n}$, and exact recovery requires the coordinate-wise rounding condition in Eq.~\eqref{eq:round-eq}. This can be restrictive in small-sample or noisy regimes: even if the estimator reduces the residual over $\mathbb{R}^{n}$, some coordinates may still fall outside the rounding radius of the true integer value.

\begin{figure*}[tbp]
    \centering
    \includegraphics[width=0.8\textwidth,
        ]{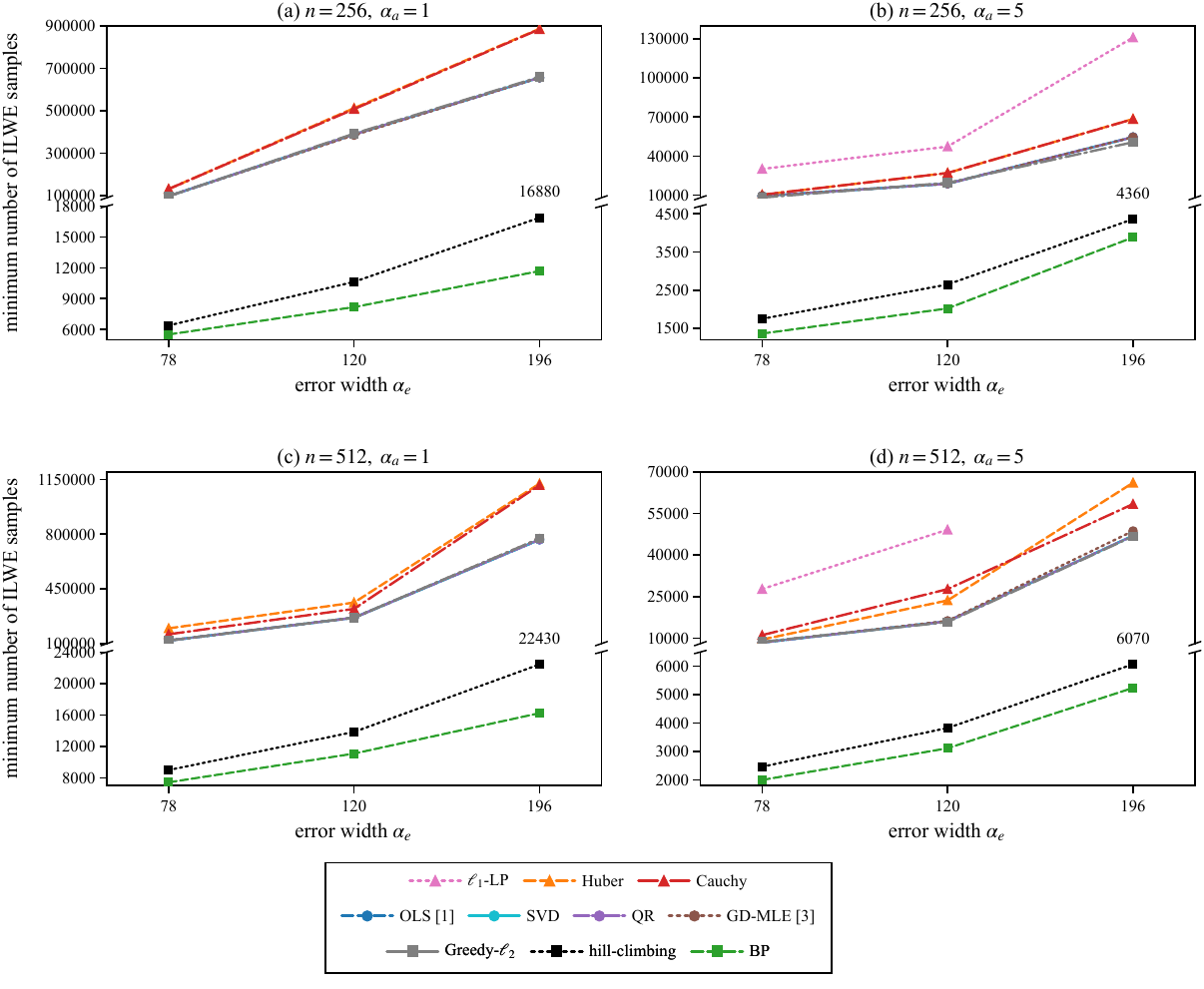}
    \caption{Average minimum samples required for solving OILWE.}
    \label{fig:ILWE}
\end{figure*}

By contrast, prior-aware discrete inference solvers search or infer directly over the bounded integer candidate space $\mathcal{S}$. Their success is governed by candidate separation rather than coordinate-wise rounding accuracy. For BP, this means that the posterior concentrates on the true secret. For objective-induced greedy, the secret should have a lower residual loss than competing integer candidates, although Greedy-$\ell_2$ may still behave similarly to least-squares regression because it ranks candidates by the $\ell_2$-residual objective. For hill-climbing, the secret should be more consistent with the effective bounded-error interval. These conditions can be summarized as
\[
\mathcal{J}(\mathbf{s}) < \mathcal{J}(\mathbf{x}),
\
\forall \mathbf{x}\in\mathcal{S},\ \mathbf{x}\neq\mathbf{s},
\]
where \(\mathcal{J}\) is the corresponding candidate cost, e.g., posterior, residual, or inconsistency cost.

This viewpoint explains why prior-aware discrete-inference solvers can reduce the finite-sample requirement. Their advantage does not come only from restricting the search space from $\mathbb{R}^n$ to $\mathcal{S}$; it also comes from exploiting structural and statistical priors, such as bounded small coefficients, non-uniform coefficient distributions, sparsity, error distributions, and effective error bounds. These priors reduce the set of plausible candidates or improve how samples distinguish them. Nevertheless, because each ILWE relation with errors, more than $n$ samples are generally still needed to separate the true secret from close competitors. Thus, the gain should be understood as a finite-sample advantage: prior-aware discrete inference can require substantially fewer samples than continuous rounding-based solvers, while still relying on enough redundant relations for reliable candidate separation.

\subsection{Experimental Results}\label{subsec3.4}

We first evaluate the solvers on ordinary ILWE as a baseline. As in prior ILWE experiments, we sample sparse secrets by setting \(\lfloor0.15n\rfloor\) entries to \(\pm1\), \(\lfloor0.15n\rfloor\) entries to \(\pm2\), and the remaining entries to \(0\). In our setting, both the public sample vectors and the error terms are sampled from uniform distributions over \([-\alpha,\alpha]\), which satisfy subgaussian properties. We test \(n\in\{256,512\}\), \(\alpha_a\in\{1,5\}\), and \(\alpha_e\in\{78,120,196\}\). For each parameter setting, the same generated sample sequence is used for all solvers. The minimum sample size is obtained by increasing the number of samples multiplicatively to locate the recovery range, and then applying binary search within this range. Recovery is declared successful only when the final rounded or decoded candidate equals the true secret.

Fig.~\ref{fig:ILWE} shows the average minimum number of OILWE samples required by different solvers in five experiments. Overall, among the three classes of recovery solvers, prior-aware discrete-inference solvers achieves the best performance. The least-squares regression solvers and Greedy-\(\ell_2\) have nearly identical thresholds, indicating they solve essentially the same least-squares problem and differ mainly in numerical implementation, and consequently their curves often overlap. Robust regression solvers such as \(\ell_1\)-LP\footnote{When the number of samples is large, the corresponding \(\ell_1\)-LP becomes large and may exceed practical time limits.}, Huber, and Cauchy offer no sample-complexity advantage in this OILWE setting, since the error is centered and uniformly distributed rather than arising from outliers or concealed samples, so downweighting large residuals does not improve recovery and may even increase the required sample size. The main reduction comes from prior-aware discrete-inference solvers, where BP consistently achieves the smallest sample requirement across all tested settings, improving over the OLS baseline by \(4.2\times\) to \(56.1\times\), and hill-climbing also benefits from the bounded small-integer prior and improves over OLS by \(3.4\times\) to \(38.9\times\), although it generally requires more samples than BP, showing that even for ordinary ILWE without leakage-specific preprocessing, exploiting the discrete prior of the secret is more important for sample efficiency than changing the numerical implementation of least-squares regression.

\section{The Fiat-Shamir Integer-LWE problem}\label{sec4}

This section studies FS-ILWE and noisy FS-ILWE induced by randomness-bit leakage in ML-DSA. We first review how the leakage models of~\cite{liu2020security,damm2025one,damm2026less} convert leaked randomness bits into FS-ILWE. We then apply the solvers introduced in the previous section to this instance and analyze their recovery conditions. Finally, we compare their sample requirements and recovery performance experimentally.

\subsection{Review of FS-ILWE}\label{subsec4.1}

We review randomness-bit leakage attacks against ML-DSA, following the line initiated by Liu et al.~\cite{liu2020security} and refined by Damm et al.~\cite{damm2025one}. These attacks convert leaky Fiat-Shamir signatures into FS-ILWE relations. Since Damm et al.~\cite{damm2025one} further identify informative relations and introduce a bit-position-independent transformation, we follow their formulation.

\paragraph{Attack setting}
We use the ML-DSA signing relation from Section~\ref{subsec:mldsa}. Fix one polynomial component of \(\mathbf{s}_1\), and denote its coefficient vector by \(\mathbf{s}\in[\pm\eta]^n\). For each released signature coefficient, the term \(\mathbf{c}\mathbf{s}_1\) induces a sparse linear relation $z_i=\langle \mathbf{c}_i,\mathbf{s}\rangle+y_i$, where \(\mathbf{c}_i\) denotes the row of the convolution matrix induced by \(\mathbf{c}\). The randomness-bit leakage attack assumes that, in addition to the public values \((\mathbf{c}_i,z_i)\), the attacker obtains one fixed bit \(y_{i,j}\) of the randomness coefficient \(y_i\). The goal is to recover the subkey \(\mathbf{s}\).

\paragraph{From leakage to FS-ILWE}
The key observation of Liu et al.~\cite{liu2020security} is that the leaked bit \(y_{i,j}\), together with \((z_i,\mathbf{c}_i)\), allows the attacker to correct the centered low-bit value \([z_i]_{2^{j-1}}\) by determining whether a carry or borrow has occurred. Let \(d_i\in\{0,1\}\) indicate whether this correction is needed. The corrected right-hand side is $\bar z_i := [z_i]_{2^{j-1}} \pm d_i2^{j-1}$. Then the attacker extracts a linear relation $\bar z_i = \langle \mathbf{c}_i,\mathbf{s}\rangle + [y_i]_{2^{j-1}}$, where $[y_i]_{2^j-1}$ denotes the centered reduction of $y_i$ modulo $2^{j-1}$. Compared with OILWE, the rows \(\mathbf{c}_i\) are sparse and structured.

\paragraph{Informative relations and effective error bound}
Damm et al.~\cite{damm2025one} observe that not all extracted relations are informative. In the high-bit regime \(2^{j-1}>\beta\), a relation is informative only if $|\bar z_i|\geq 2^{j-1}-\beta$. For each retained relation, the shift $\tilde z_i = \bar z_i-\operatorname{sign}(\bar z_i)(2^{j-1}-\beta)$ gives $\tilde z_i=\langle \mathbf{c}_i,\mathbf{s}\rangle+\tilde y_i,\ |\tilde y_i|\leq \beta$. Thus, after filtering and shifting, the error bound no longer grows with the leakage-bit position \(j\). In the low-bit regime \(2^{j-1}\leq\beta\), all relations are already informative and no shift is needed. Both cases are summarized by a effective error bound $\beta_{\mathrm{eff}}=\min\{\beta,2^{j-1}\}$. After preprocessing, the attacker obtains FS-ILWE instance $\tilde{\mathbf{z}}=\mathbf{C}\mathbf{s}+\tilde{\mathbf{y}},\ |\tilde{\mathbf{y}}|_{\infty}\leq\beta_{\mathrm{eff}}$.

\paragraph{Secret recovery under exact leakage}
Under exact leakage, prior works~\cite{liu2020security,damm2025one,damm2026less} solve the FS-ILWE instance by using least-squares regression solvers followed by coordinate-wise rounding. However, this continuous-regression-then-rounding solvers may require many informative relations, especially for higher leakage positions or larger ML-DSA parameter sets. Schubert et al.~\cite{schubert2026descent} instead formulate recovery as a bounded-error local-search problem over integer subkeys, showing that exploiting the small-integer prior can reduce the informative-relation requirement.

\paragraph{Noisy randomness-bit leakage}
In practice, side-channel extraction may produce an incorrect estimate of the randomness leakage bit. Following the noisy leakage analysis of~\cite{damm2026less}, we write the observed leakage as $y_{i,j}^{\mathrm{obs}} = y_{i,j}\oplus\nu_i,\ \nu_i\sim\mathrm{Ber}(p_{\mathrm{flip}})$, where \(p_{\mathrm{flip}}<1/2\) denotes the independent bit-flip probability. The observed bit is then processed using the same filter-and-shift relation-extraction procedure as in~\cite{damm2026less}. If \(\nu_i=0\), preprocessing yields a FS-ILWE instance. Otherwise, the incorrect leakage bit may alter the centered reduction, the informative-relation decision, and the shift applied during preprocessing, resulting in an incorrectly transformed relation.

Consequently, noisy leakage does not simply increase the effective error bound. Instead, the extracted instance consists of a mixture of valid bounded-error relations and incorrectly transformed relations induced by bit flips, as characterized in~\cite{damm2026less}. From the perspective of secret recovery, these incorrectly transformed relations behave as unreliable constraints. Least-square regression solvers are particularly sensitive to their influence because all relations contribute equally to the $\ell_2$-loss objective. Robust regression solvers mitigate this effect through robust loss functions, whereas prior-aware discrete-inference solvers search directly over the bounded integer secret space and identifies the candidate secret that is most consistent with the extracted relation set.

\subsection{Solving FS-ILWE Instance} \label{subsec4.2}

After leakage extraction, informative-relation filtering, and the bit-position-independent transformation reviewed above, the attacker obtains an FS-ILWE instance $\tilde{\mathbf{z}}=\mathbf{C}\mathbf{s}+ \tilde{\mathbf{y}}$. For least-squares regression solvers and robust regression solvers, we use the same estimators as in Sec.~\ref{subsec3.1} and Sec.~\ref{subsec3.2}, but with the input \((\mathbf{C},\tilde{\mathbf{z}})\). The least-squares solvers minimize \(\|\mathbf{C}\mathbf{x}-\tilde{\mathbf{z}}\|_2^2\) over \(\mathbb{R}^n\) and then apply coordinate-wise rounding, while robust solvers replace the $\ell_2$ loss with an \(\ell_1\) loss, Huber loss, or Cauchy loss. 

For prior-aware discrete-inference solvers, FS-ILWE satisfies the BP recovery condition because each transformed relation can still be written as a sparse likelihood factor over the same bounded secret space. Specifically, the secret satisfies \(\mathbf{s}\in\mathcal{S}\subseteq\{-\eta,\ldots,\eta\}^n\), and its coefficient-wise distribution can be used as a prior when available. Each row of \(\mathbf{C}\) contains only \(\tau\) nonzero entries in \(\{\pm1\}\), so every relation involves only a small subset of secret coefficients. After the bit-position-independent transformation, the residual is controlled by the bounded-error model with scale \(\beta_{\mathrm{eff}}\). Thus, BP combines the small-secret prior with likelihood factors induced by these sparse bounded-error relations. When the accumulated likelihoods make the true integer secret clearly separated from competing candidates, BP can recover $\mathbf{s}$ without relying on a real-valued estimate followed by rounding.

The same structure also explains when BP fails in FS-ILWE. If the transformed relations become too noisy, the likelihood factors lose discriminative power: the likelihood gap between the best-supported and worst-supported candidate values becomes small, and the messages passed in BP become close to uniform. Equivalently, each relation carries little information for distinguishing the true secret from nearby wrong integer candidates. In this case, the posterior marginals cannot separate from the prior, or may separate toward an incorrect candidate, so BP cannot reliably recover the secret. This situation may occur under high bit-flip noise, severe relation corruption, or lower-bit model mismatch where the assumed bounded-error FS-ILWE likelihood no longer matches the actual residual distribution.

Objective-induced greedy and hill-climbing use the same structural priors in a different way. Greedy searches over \(\mathcal{S}\) by selecting updates that decrease a residual-induced objective, while hill-climbing evaluates whether candidate residuals are consistent with the interval \([-\beta_{\mathrm{eff}},\beta_{\mathrm{eff}}]\). Therefore, prior-aware discrete-inference solvers exploit not only the finite candidate space, but also the sparse challenge structure and bounded-error model, allowing them to compare integer candidates directly instead of relying on coordinate-wise rounding.

\vspace{-0.2cm}

\subsection{Sample-Complexity Interpretation for FS-ILWE} \label{subsec4.3} 

The least-squares informative-relation requirement for FS-ILWE has been analyzed in prior work~\cite{liu2020security,damm2025one,damm2026less}. After valid relation extraction, the problem becomes a bounded-error FS-ILWE instance with effective error bound \(\beta_{\mathrm{eff}}\). In this setting, OLS followed by coordinate-wise rounding succeeds with probability at least \(1-\mu\) once
\begin{equation}
    m_{\mathrm{inf}}\geq \frac{2n}{3\tau} \left((2\beta_{\mathrm{eff}}+1)^2-1\right) \ln\left(\frac{2n}{\mu}\right).
    \label{eq:FS-ILWE-minsample}
\end{equation}
Thus, the OLS requirement decreases with the effective error bound, provided that the extracted relations are valid bounded-error FS-ILWE samples.

The validity of \(\beta_{\mathrm{eff}}\) depends on the leakage position. Let \(\beta\) be the worst-case bound on the secret-dependent term. When \(2^{j-1}>\beta\), the leaked-bit scale dominates the worst-case shift, filter-and-shift preprocessing gives \(\beta_{\mathrm{eff}}=\beta\), and the bound above applies directly. When \(2^{j-1}\leq\beta\), the nominal bound becomes \(\beta_{\mathrm{eff}}=2^{j-1}\), but this does not mean that \(j\) can be decreased arbitrarily. The worst-case bound is conservative for ML-DSA because \(\langle \mathbf{c},\mathbf{s}\rangle\) is concentrated around zero, with variance $\sigma^2=\frac{(2\eta+1)^2-1}{12}\cdot \tau$. This explains why practical attacks may start around \(j=6\) for ML-DSA-44 and ML-DSA-87, and around \(j=7\) for ML-DSA-65. If the leakage position is pushed further downward, the public value \(z\) may be assigned to a wrong centered interval or preprocessing branch. The resulting error is no longer simply a smaller bounded error, but may be correlated with \(\langle \mathbf{c},\mathbf{s}\rangle\), causing biased extracted relations. Earlier lower-bit analyses in~\cite{wang2021integer} describe this mismatch as an incorrectly inferred carry or borrow value, in which case least squares may recover a scaled vector \(\lambda s\) and require additional correction. Here, \(\lambda\) is the attenuation factor induced by imperfect carry/borrow inference. Such correction is only manageable for moderately low leakage positions.

In the noisy leakage setting, each leaked bit may be flipped independently with probability \(p_{\mathrm{flip}}<1/2\). If the transformed noisy FS-ILWE instance remains valid, the least-squares estimate is attenuated approximately by a factor \(1-2p_{\mathrm{flip}}\), so OLS applies the correction \(\hat{s}\mapsto \hat{s}/(1-2p_{\mathrm{flip}})\) before rounding. The sample requirement is then increased by both the rescaling factor and the extra variance from flipped relations. The noisy-leakage analysis of~\cite{damm2026less} gives the following
sufficient asymptotic condition in the high-bit regime:
\begin{equation}
    \begin{aligned}
    m_{\mathrm{inf}}
    &\ge
    8\ln\!\left(\frac{2n}{\mu}\right)
    \frac{1}{(1-2p_{\mathrm{flip}})^2}  \\
    &\quad \cdot
    \left[
    \frac{n}{\tau}
    \left(
    \frac{\beta(\beta+1)}{3}
    +2p_{\mathrm{flip}}\beta^2
    \right)
    +2p_{\mathrm{flip}}\|\mathbf{s}\|_2^2
    \right].
    \end{aligned}
\end{equation}
This highlights that noisy leakage increases the OLS requirement mainly through \((1-2p_{\mathrm{flip}})^{-2}\) and the additional corrupted-relation variance. The bound should be interpreted only when the bit-flip and transformed-relation instance are valid, not in very low-bit regimes where extraction itself becomes structurally mismatched.

\begin{figure*}[tbp]
    \centering
    \includegraphics[
        width=0.9\textwidth,
        keepaspectratio
    ]{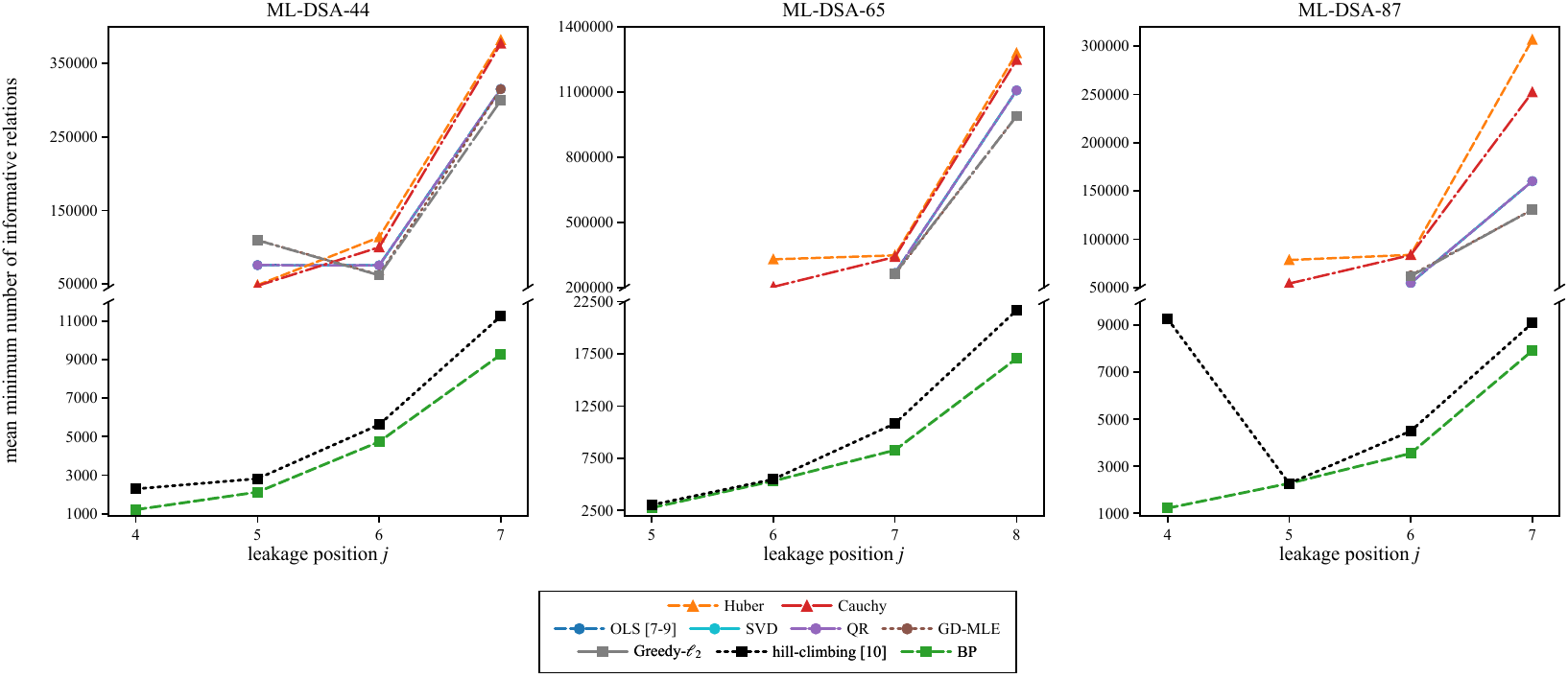}
    \caption{Average minimum informative relations required for solving FS-ILWE. Curves stop when the solver fails to recover within the tested relation bound.}
    \label{fig:fsilwe_noiseless_all_solvers}
\end{figure*}

Robust regression solvers should be viewed as a corrupted-relation baseline. In FS-ILWE, the transformed samples already satisfy a bounded-error model, so \(\ell_1\), Huber, and Cauchy losses are not expected to improve the OLS sample threshold; they still output real-valued estimates and require coordinate-wise rounding. Their advantage appears when some relations become unreliable, either because low leakage positions introduce branch errors correlated with \(\langle \mathbf{c}, \mathbf{s}\rangle\), or because bit flips occur. Robust losses can help when such unreliable relations appear as large-residual corruptions, but they may not outperform bias-corrected OLS when the noise mainly produces structured bias or global signal attenuation.

Prior-aware discrete-inference solvers follow a different principle. Their success depends on candidate separation rather than coordinate-wise rounding. In FS-ILWE, BP, objective-induced greedy, and hill-climbing compare bounded integer candidates using the small-secret prior, sparse challenge rows, and bounded residual information. BP succeeds when the posterior concentrates on the true secret; hill-climbing succeeds when the true candidate is more consistent with the effective bounded-error interval. Greedy-\(\ell_2\), however, should be distinguished from BP and hill-climbing: although it searches over bounded integers, its score is still based on $\ell_2$ residuals loss, so it can behave similarly to least-squares regression when that objective dominates.

At lower leakage positions or under noisy leakage, extracted relations may become partially bounded-error or corrupted. Prior-aware solvers can still benefit from the finite secret space and sparse challenge structure, but their likelihoods or consistency scores may become mismatched if they assume a bounded-error model. When flip rates or sample reliability can be estimated, this information can be incorporated through noise-aware likelihoods, weighted residual scores, or branch-aware consistency tests. Therefore, their advantage comes not only from restricting the search to the finite secret space, but also from using structural priors about valid relations, corrupted relations, and sample reliability.

\vspace{-0.2cm}

\subsection{Experimental Results}\label{subsec4.4} 

\subsubsection{Experimental Results on FS-ILWE} \label{subsec:fsilwe_noiseless_experiments}

We evaluate the solvers on FS-ILWE instances derived from single-bit randomness leakage in ML-DSA. For each signature, we use the public challenge row, the response coefficient, and one leaked bit of the randomness to construct FS-ILWE relations via leakage extraction and filter-and-shift preprocessing. As in prior work, we report the number of informative relations rather than signatures, since the latter depends on the probability of yielding an informative relation. We test all three ML-DSA parameter sets: for ML-DSA-44 and ML-DSA-87, the leakage positions are 4, 5, 6, and 7; for ML-DSA-65, the leakage positions are 5, 6, 7, and 8, covering both statistically meaningful and lower-bit boundary regimes. For each setting, we run five seeds and report the average minimum number of informative relations. All solvers use the same relation sequence. The minimum is found by first scaling the number of relations multiplicatively to locate the recovery range, then applying binary search. Recovery is declared only when all 256 coefficients of the target subkey are recovered.

\begin{figure}[!t]
    \centering
    \includegraphics[
        width=\columnwidth]{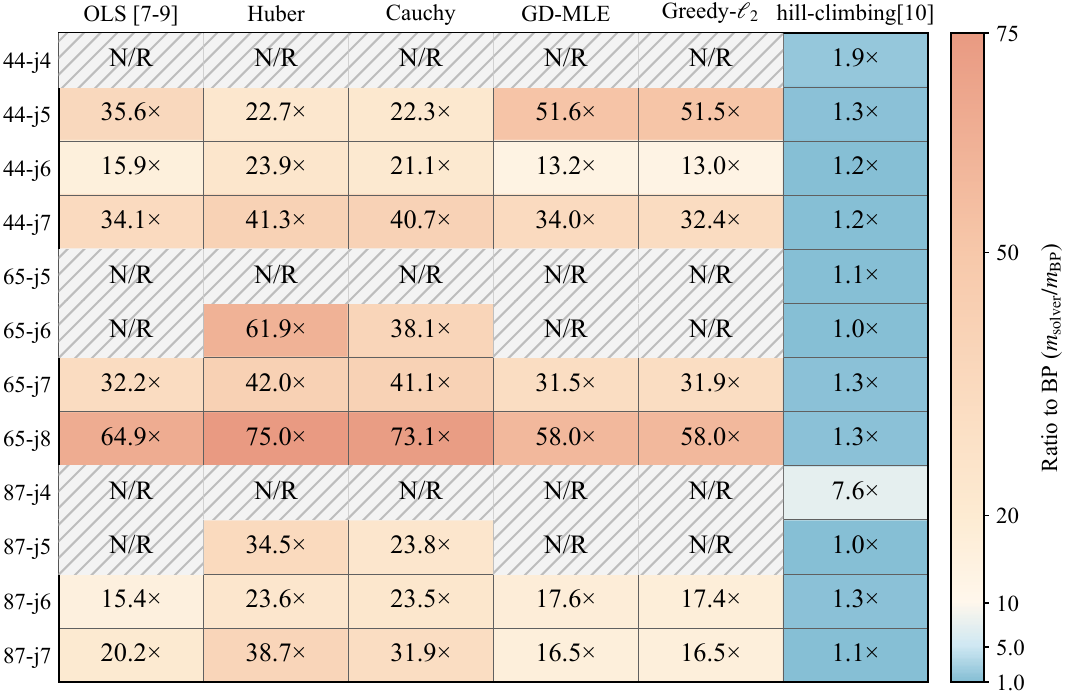}
    \caption{Relative informative-relation requirements over BP on solving FS-ILWE. Hatched gray cells indicate unrecovered cases.}
    \label{fig:fsilwe_noiseless_bp_ratio}
\end{figure}

Fig.~\ref{fig:fsilwe_noiseless_all_solvers} shows the average minimum number of informative relations required by each solver, and Fig.~\ref{fig:fsilwe_noiseless_bp_ratio} shows the corresponding ratio to BP. The least-squares regression solvers and Greedy-\(\ell_2\) differ only in numerical solution strategies rather than objectives, leading to nearly identical thresholds that overlap in the plots. Robust regression solvers are not consistently better than OLS, but they can recover some lower-bit boundary cases where OLS fails. This is consistent with the previous analysis: when the leakage position is too low, the extracted relations may deviate from the FS-ILWE, and reducing the influence of unreliable relations can sometimes improve stability.

The dominant improvement comes from prior-aware discrete-inference solvers. BP consistently achieves the smallest informative-relation requirement across almost all tested settings, typically needing only a few hundred to about \(1.7\times10^4\) relations, and reducing the requirement by \(15.4\times\) to \(64.9\times\) compared to OLS. Hill-climbing, as a discrete local-search refinement over OLS, already improves over least-squares solvers, but BP further reduces the required relations by up to \(7.6\times\). Moreover, prior-aware discrete-inference solvers recover at lower leakage positions than robust regression solvers: while Huber and Cauchy occasionally succeed where OLS fails, BP and hill-climbing consistently recover in more challenging settings. This aligns with earlier observations that when extracted relations deviate from the FS-ILWE, solver choice becomes critical, and exploiting the discrete prior significantly improves robustness. Overall, these results confirm that prior-aware discrete inference is the most sample-efficient approach, outperforming both continuous regression and robust methods, especially in low-leakage regimes.

\subsubsection{Experimental Results on Noisy FS-ILWE}\label{subsec:fsilwe_noisy_experiments}

\begin{figure*}[!htbp]
    \centering
    \includegraphics[
        width=0.9\textwidth,
        keepaspectratio
    ]{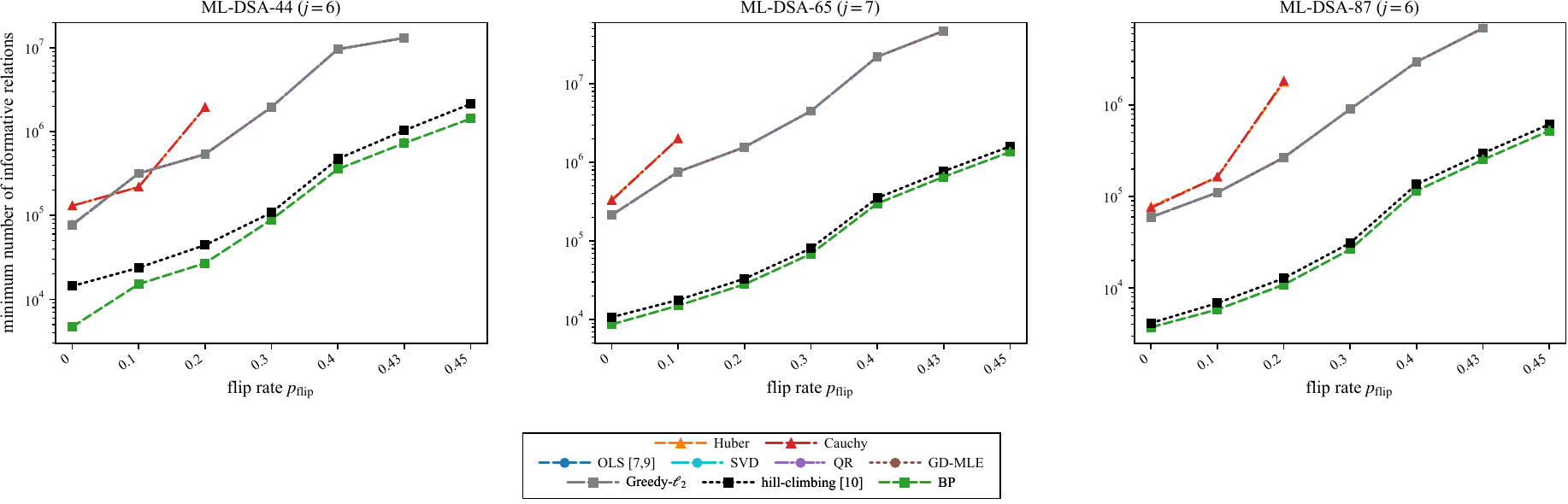}
    \caption{Average minimum informative relations required for solving noisy FS-ILWE. Curves stop when the solver fails to recover within the tested relation bound.}
    \label{fig:solver_comparison_noisy}
\end{figure*}

We further evaluate the solvers under noisy FS-ILWE, where each leaked randomness bit is independently flipped with probability \(p_{\mathrm{flip}}\). Prior OLS-based work \cite{damm2026less} shows that recovery remains possible under this bit-flip model with filter-and-shift preprocessing and bias correction, reaching flip rates up to \(p_{\mathrm{flip}}=0.43\). The hill-climbing attack \cite{schubert2026descent} further preserves this noisy leakage setting and demonstrates recovery at \(p_{\mathrm{flip}}=0.45\). In our experiments, we use the same ML-DSA parameter sets and leakage-position as in the noise-free setting. We test several flip rates \(p_{\mathrm{flip}}\in[0,0.45]\).

Fig.~\ref{fig:solver_comparison_noisy} shows that the required number of informative relations increases rapidly as the flip rate grows: least-squares regression solvers and Greedy-\(\ell_2\) become increasingly data-hungry and often require millions of relations or fail at high flip rate, while robust variants do not consistently improve performance because the sample filtering step removes extreme residual errors, leaving few large outliers for robust loss functions to mitigate. In contrast, prior-aware discrete inference remains effective. Compared with least-squares solvers, BP reduces the informative-relation requirement by \(10.5\times\) to \(73.9\times\). It also consistently requires fewer informative relations than hill-climbing.

\section{The Concealed Integer-LWE problem}\label{sec5}
This section studies profiling-induced CILWE instance in ML-DSA, following Ulitzsch et al.~\cite{ulitzsch2024profiling}, Berzati et al.~\cite{berzati2023exploiting}, and Damm et al.~\cite{damm2025solving}. We briefly review how profiling leakage yields mixtures of informative and concealed samples, and then apply the solver framework of Section~\ref{sec3} to compare recovery behavior and sample requirements under different concealment rates.

\subsection{Review of the CILWE}\label{subsec5.1}

We review the Concealed ILWE (CILWE) of Damm et al.~\cite{damm2025solving} and its connection to profiling side-channel attacks on ML-DSA~\cite{ulitzsch2024profiling,berzati2023exploiting}. Unlike FS-ILWE, which is induced by randomness-bit leakage, CILWE arises when profiling leakage is used to select signature samples whose randomness coefficient is predicted to be zero. It provides a second ILWE-family instance in which the selected samples mix informative zero-error samples with concealed samples.

\paragraph{From zero-value leakage to CILWE samples}
We use the coefficient-level ML-DSA signing relation from Section~\ref{subsec:mldsa}. For one polynomial component $\mathbf{s}$ of $\mathbf{s}_1$, each selected coefficient satisfies $z_i=\langle \mathbf{c}_i,\mathbf{s}\rangle+y_i$. Profiling attacks aim to select samples with $y_i=0$. A correct prediction gives an exact equation $z_i=\langle \mathbf{c}_i,\mathbf{s}\rangle$, whereas a false positive yields a concealed sample with $y_i\neq0$. Thus, the collected samples consists of informative zero-error samples with probability $1-p_{\mathrm{con}}$ and concealed samples with probability $p_{\mathrm{con}}$:
\begin{equation}
    \begin{aligned}
    \mathbf{z} &= \mathbf{C}\mathbf{s}+\mathbf{y},\ 
    y_i =
    \begin{cases}
    0, & \text{with } 1-p_{\mathrm{con}},\\
    \tilde{y}_i\leftarrow\chi_{\mathrm{con}}, & \text{with } p_{\mathrm{con}},
    \end{cases}
    \end{aligned}
    \label{eq:CILWE_ML-DSA}
\end{equation}
where $p_{\mathrm{con}}$ is the concealment rate.

\paragraph{Difference from ordinary ILWE}
CILWE mainly differs from ordinary ILWE in its error structure. In ordinary ILWE, the error is modeled as centered error independent of the public sample matrix and the secret key. In CILWE, concealed samples come from false-positive profiling decisions or zero-knowledge signing outputs, and therefore do not behave as reliable small errors. When mixed with zero-error samples, they act as corrupted relations or outliers. This makes ordinary least squares unstable and motivates robust regression solvers, such as Huber and Cauchy regression, as well as prior-aware discrete-inference solvers that exploits the bounded integer structure of the secret key.

\vspace{-0.2cm}

\subsection{Solving CILWE Instance}\label{subsec5.2}

Given the selected samples in~\eqref{eq:CILWE_ML-DSA}, the recovery task is to recover the bounded integer subkey \(\mathbf{s}\) from the mixed system $\mathbf{z}=\langle\mathbf{C},\mathbf{s}\rangle+\mathbf{y}$. Unlike ordinary ILWE, the error term has a mixture structure. Informative samples satisfy \(y_i=0\) and give exact constraints on \(\mathbf{s}\), whereas concealed samples arise from false-positive profiling decisions and may behave as zero-knowledge samples, corrupted relations, or large-residual outliers. Thus, only about \(m_{\mathrm{eff}}=(1-p_{\mathrm{con}})m\) selected samples provide reliable zero-error information.

For regression solvers, we use the same estimator forms as in Sec.~\ref{subsec3.1} and Sec.~\ref{subsec3.2}, but with the CILWE input \((\mathbf{C},\mathbf{z})\). For a candidate \(\mathbf{x}\), the residual is $r_i(\mathbf{x})$. Least-squares regression solvers minimize squared residuals over \(\mathbb{R}^n\) and then apply coordinate-wise rounding. Since concealed samples do not behave as centered independent small errors, they can strongly influence the squared-loss estimate. Robust regression solvers replace the squared loss with an \(\ell_1\), Huber, or Cauchy loss, thereby reducing the effect of concealed samples in the continuous objective.

For BP, the input consists of \((\mathbf{C},\mathbf{z})\), the bounded coefficient space \(\mathcal{S}_j\) for each secret coordinate, the coefficient-wise secret prior, and a CILWE instance determined by the concealment rate \(p_{\mathrm{con}}\). CILWE satisfies the BP recovery condition in~\ref{subsec3.3} because each sample still induces a sparse likelihood factor over the same bounded integer secret space. Let \(\mathcal{N}(i)\) be the support of the \(i\)-th sparse challenge row. For a candidate assignment \(\mathbf{x}_{\mathcal{N}(i)}\), the residual is $r_i(\mathbf{x}) = z_i-\sum_{j\in\mathcal{N}(i)} c_{i,j}x_j$. In the ideal zero-error CILWE, informative samples strongly support assignments with \(r_i(\mathbf{x})=0\), while concealed samples should not be treated as reliable constraints. Therefore, we use a mixture likelihood factor
\begin{equation}
    \phi_i(\mathbf{x}_{\mathcal{N}(i)})
    =
    (1-p_{\mathrm{con}})D_0(r_i(\mathbf{x}))
    +
    p_{\mathrm{con}}D_{\mathrm{con}}(r_i(\mathbf{x})),
    \label{eq:BP_CILWE}
\end{equation}
where \(D_0\) is concentrated at zero, e.g., \(D_0(r)=\mathbf{1}[r=0]\) in the exact setting, and \(D_{\mathrm{con}}\) is a broad or nearly flat distribution describing concealed samples. If informative samples contain small independent errors, \(D_0\) can be replaced by a narrow error distribution centered at zero.

Under this model, BP applies the same message-passing updates as in Sec.~\ref{subsec3.3}, but with the CILWE-specific likelihood factors in Eq.~\eqref{eq:BP_CILWE}. When \(p_{\mathrm{con}}\) is moderate, the informative component \((1-p_{\mathrm{con}})D_0\) still creates a clear likelihood gap between candidate assignments that satisfy the zero-error relation and those that do not. As these sparse constraints accumulate over many samples, the posterior mass of the true bounded integer secret can separate from that of wrong candidates, and BP can recover the secret key.

However, when \(p_{\mathrm{con}}\) becomes large, CILWE no longer provides enough informative constraints for BP. In this regime, the mixture likelihood is dominated by \(D_{\mathrm{con}}\), and the factor values assigned to different candidate assignments become nearly indistinguishable. Equivalently, the gap between the maximum and minimum likelihood values of a sample becomes very small, so each message carries little information about the true secret. The resulting marginals remain close to the prior, or separate only weakly and unreliably. Therefore, when the concealment rate is too high, the posterior-separation condition of BP is not satisfied, and BP can no longer reliably solve the CILWE instance.

For objective-induced greedy, the solver keeps an integer candidate \(\mathbf{x}^{(t)}\in\mathcal{S}\) and selects local updates according to a CILWE-aware cost, for example the negative log-likelihood $\mathcal{L}_{\mathrm{CILWE}}(\mathbf{x}) =
\sum_{i=1}^{m} -\log ( (1-p_{\mathrm{con}})D_0(r_i(\mathbf{x}))+$ $  p_{\mathrm{con}}D_{\mathrm{con}}(r_i(\mathbf{x})))$. Equivalently, one may use a robust surrogate such as an \(\ell_1\), Huber, Cauchy, truncated, or consistency-based loss. The greedy update score is the decrease of this cost after a local integer modification. Thus, zero-error samples reward candidates satisfying many exact constraints, while concealed samples are downweighted or capped.

Hill-climbing can be viewed as a consistency-based local search that scores candidates by the number or magnitude of violated zero-error constraints, possibly with a tolerance for small independent errors. Hence, in CILWE, prior-aware solvers exploit the bounded integer secret, sparse challenge rows, zero-error samples, concealment rate, and concealed-sample distribution, and compare integer candidates directly.

\subsection{Sample-Complexity Interpretation for CILWE}\label{subsec5.3}

We first consider the ideal case \(p_{\mathrm{con}}=0\). In this case, every selected sample is zero-error, and CILWE reduces to an exact sparse linear system. If the challenge-derived matrix \(\mathbf{C}\) has full column rank, the secret can be recovered from sufficiently many independent samples. Thus, the basic information requirement is $m(0)\approx n$, up to the additional samples needed to ensure that the sparse challenge rows span the \(n\)-dimensional secret space. In this ideal regime, all solvers have essentially the same information source, because all selected samples provide exact constraints.

When \(p_{\mathrm{con}}>0\), only a fraction \(1-p_{\mathrm{con}}\) of the selected samples are informative. Therefore, a first necessary condition is
\begin{equation}
    m(p_{\mathrm{con}}) \gtrsim \frac{n}{1-p_{\mathrm{con}}}
    \label{eq:CILWE_p}
\end{equation}
This is only a counting lower bound and ignores sparse-row coverage, rank, and concealed-sample interference. It does not fully determine the actual sample requirement, because the concealed samples are still present in the system and may behave as zero-knowledge samples, corrupted relations, or large-residual outliers.

For least-squares regression solvers, the main difficulty is the rounding margin. Least squares and GD-MLE still produce a real-valued estimate and require the coordinate-wise rounding condition. However, concealed samples do not provide centered independent small errors; they can bias the real-valued estimate or create large residuals that dominate the squared-loss objective. As the concealment rate \(p_{\mathrm{con}}\) increases, the pre-rounding error is more likely to exceed \(1/2\) in at least one coordinate, causing rounding failure even when many zero-error samples are available.

Robust regression solvers have the same final rounding requirement, but changes how concealed samples affect the estimate. In the IRLS view, large-residual concealed samples receive smaller weights, so their contribution to the pre-rounding error is reduced. This explains why Huber, and Cauchy regression can tolerate larger concealment rates than ordinary least squares. Huber keeps quadratic curvature near zero-error samples while clipping large residuals, whereas Cauchy suppresses very large residuals more aggressively but leads to a non-convex objective.

Prior-aware discrete-inference solvers follow a different sample-complexity principle. They do not require a real-valued estimate to lie inside a rounding interval. Instead, they compare bounded integer candidates using the secret prior, sparse challenge rows, exact zero-error samples, and the concealment model. For BP, the mixture factor in~\eqref{eq:BP_CILWE} allows zero-error samples to strongly support assignments satisfying \(r_i(\mathbf{x})=0\), while concealed samples contribute only weak or low-confidence evidence. For greedy, a CILWE-aware loss or its robust surrogate downweights or caps concealed residuals when scoring local integer updates. In contrast, Greedy-\(\ell_2\) does not fully exploit the CILWE mixture structure: although it searches over the bounded integer space, its squared-residual score remains strongly affected by zero-knowledge samples, so it can degrade similarly to OLS as the concealment rate increases. For hill-climbing, the score can be based on the number or magnitude of violated zero-error constraints, with tolerance if small independent errors are allowed.

Thus, the relevant condition for prior-aware discrete-inference solvers is candidate separation: the true integer secret must be better supported than nearby wrong candidates. The lower bound in~\eqref{eq:CILWE_p} indicates how many zero-error samples are needed in principle, but additional samples may be required to overcome concealed samples and local ambiguities. Nevertheless, because BP, greedy, and hill-climbing can use the mixture structure of CILWE rather than treating all samples as ordinary regression points, they can reduce the finite-sample requirement compared with regression estimation followed by rounding, especially when the concealment rate is moderate and the zero-error samples remain sufficiently informative.

\begin{figure}[tbp]
    \centering
    \includegraphics[width=\columnwidth]{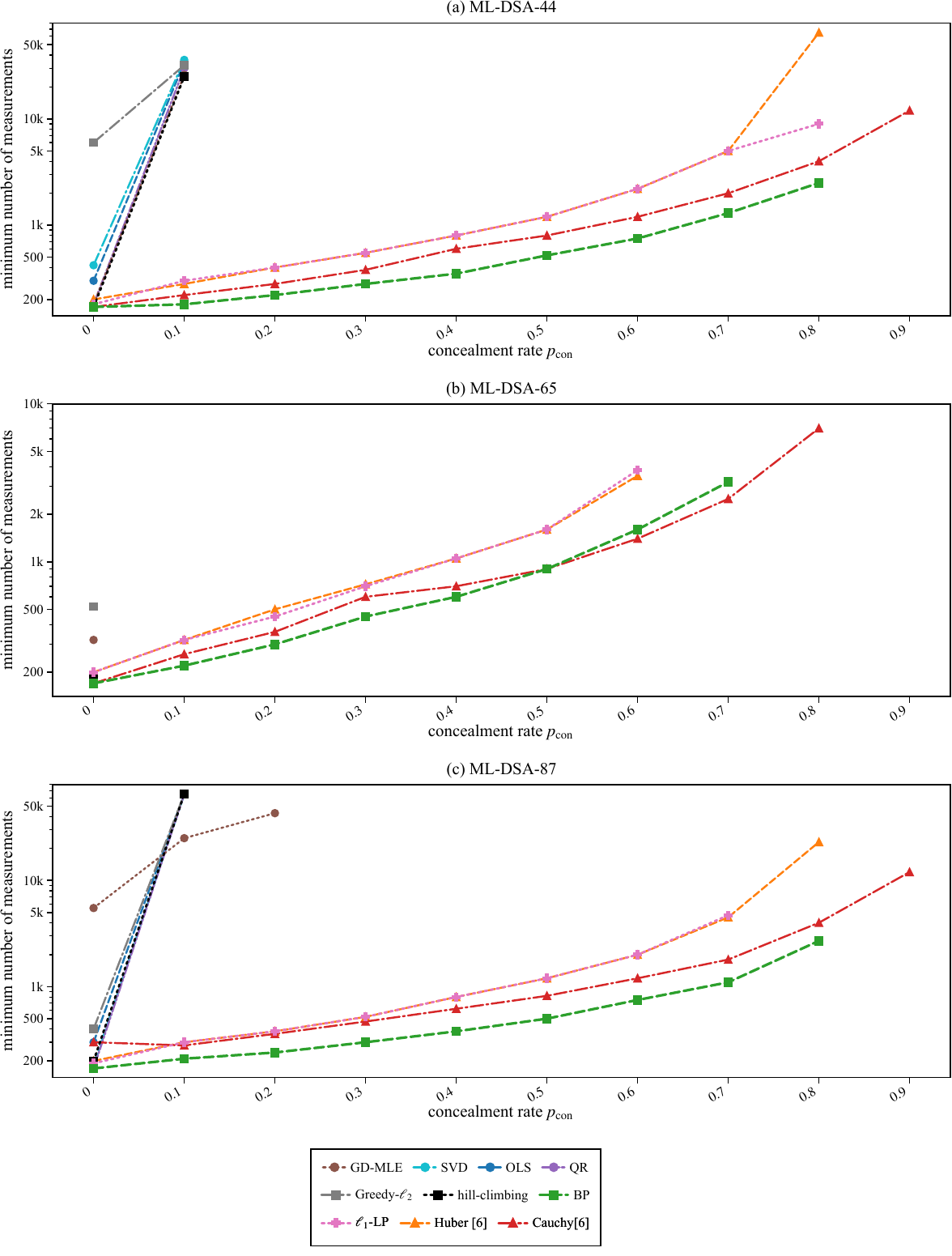}
    \caption{Average minimum samples required for solving CILWE. Curves stop when the solver fails to recover within the tested sample bound.}
    \label{fig:cilwe_solver_comparison}
\end{figure}

\subsection{Experimental Results}\label{subsec5.4}

We finally evaluate the solvers on Concealed ILWE instances. Following the simulation setting of prior CILWE work, each instance is generated as a mixture of informative and concealed samples. For each ML-DSA parameter set, we sample a 256-dimensional secret subkey \(\mathbf{s}\) from the corresponding small-secret range \([-\eta,\eta]\). Each generated measurement is a zero-error sample with probability \(1-p_{\mathrm{con}}\), and a zero-knowledge concealed sample with probability \(p_{\mathrm{con}}\), where \(p_{\mathrm{con}}\) is the concealment rate. We evaluate ML-DSA-44, ML-DSA-65, and ML-DSA-87 under different concealment rates \(p_{\mathrm{con}}\in[0,0.9]\). For each parameter setting, all solvers are tested on the same generated measurement sequence, and we record the minimum number of samples required to recover the complete 256-dimensional subkey.

Fig.~\ref{fig:cilwe_solver_comparison} shows the minimum number of samples required by different solvers as the concealment rate increases. Least-squares regression solvers and Greedy-\(\ell_2\) degrade quickly and only work at very low concealment rates, as zero-knowledge samples violate independence assumptions and bias regression. Hill-climbing fails because its empirically set effective boundary becomes unreliable as the concealment rate increases. In contrast, robust regression solvers \(\ell_1\)-LP, Huber, and especially Cauchy, perform better by reducing the impact of large-residual concealed samples.

BP achieves the lowest measurement requirement at low and moderate concealment rates $(p_{\mathrm{con}<0.9})$ across all ML-DSA parameter sets, benefiting from the small-secret prior and discrete candidate comparison. However, at high concealment rates $(p_{\mathrm{con}\leq 0.9})$, Cauchy regression remains recoverable at larger \(p\) values than BP in the tested setting. Thus, unlike OILWE and FS-ILWE, CILWE shows a clearer trade-off: BP is most efficient at low-to-moderate concealment, while Cauchy is more robust at high concealment rates.

\section{Conclusion}\label{sec6}

In this paper, we studied the solver stage of randomness leakage attacks against ML-DSA through a unified framework. The framework covers OILWE, FS-ILWE, and CILWE, and compares least-squares regression solvers, robust regression solvers, and prior-aware discrete-inference solvers under the same ILWE-type problem formulation.

Our results show that the choice of solver can substantially affect the efficiency of leakage attacks. Least-squares and robust regression solvers rely on real-valued estimation followed by coordinate-wise rounding, whereas prior-aware discrete-inference solvers exploit the bounded small-integer secret space directly. Experiments on OILWE, FS-ILWE, noisy FS-ILWE, and CILWE show that prior-aware discrete-inference solvers can reduce sample requirements in several regimes, while robust regression remains useful when corrupted or concealed samples dominate.

\bibliographystyle{IEEEtran}
\bibliography{main_tifs_zhangph}

\end{document}